\documentclass[final]{cjs}
\usepackage{amssymb}
\usepackage{subfigure, graphicx}

\title{On non-existence of a one factor interest rate model for volatility averaged  generalized Fong--Vasicek term structures}

\author{Be\'ata Stehl\'ikov\'a\, and\, Daniel \v{S}ev\v{c}ovi\v{c}$^1$}
\begin{document}

%% Start and end pages of the document. Do not chenge
\CJSLogo{1}{8}

\maketitle
\footnotetext[1]{Dept.of Appl. Math. and Statistics, Comenius University Bratislava, 842 48 Bratislava, Slovakia}

\begin{abstract}
We study the generalized Fong--Vasicek two-factor interest rate model with stochastic volatility. In this model dispersion is assumed to follow a non-negative process with volatility proportional to the square root of dispersion, while the drift is assumed to be a general function. We consider averaged bond prices with respect to the limiting distribution of stochastic dispersion. The averaged bond prices depend on time and current level of the short rate like it is the case in many popular one-factor interest rate model including, in particular, the Vasicek and Cox--Ingersoll-Ross model. However, as a main result of this paper we show that there is no such one-factor model yielding the same bond prices as the averaged values described above.
\end{abstract}



\pagestyle{myheadings}
\thispagestyle{plain}
\markboth{B. STEHL\'IKOV\'A and D. \v{S}EV\v{C}OVI\v{C} }{On non-existence of a one factor interest rate model for volatility averaged term structures}

\section{Overview of term structure modeling}

The purpose of this paper is to study the generalized Fong--Vasicek two-factor interest rate model with stochastic volatility. In this model the dispersion of the stochastic short rate (square of volatility)  is assumed to be stochastic as well and it follows a non-negative process with volatility proportional to the square root of dispersion. The drift of the stochastic process for the dispersion is assumed to be in a rather  general form including, in particular, linear function having one root (yielding the original Fong--Vasicek model, cf. \cite{fv}) or a cubic like function having three roots (yielding a generalized Fong--Vasicek model for description of the volatility clustering, see e.g. \cite{iscam05}). We consider averaged bond prices with respect to the limiting distribution of stochastic dispersion. The averaged bond prices depend on time and current level of the short rate like it is the case in many popular one-factor interest rate model including in particular the Vasicek and Cox--Ingersoll-Ross model. However, as a main result of this paper we show that there is no such one-factor model yielding the same bond prices as the averaged values described above.

\begin{figure}
\begin{center}
\includegraphics[width=0.9\textwidth]{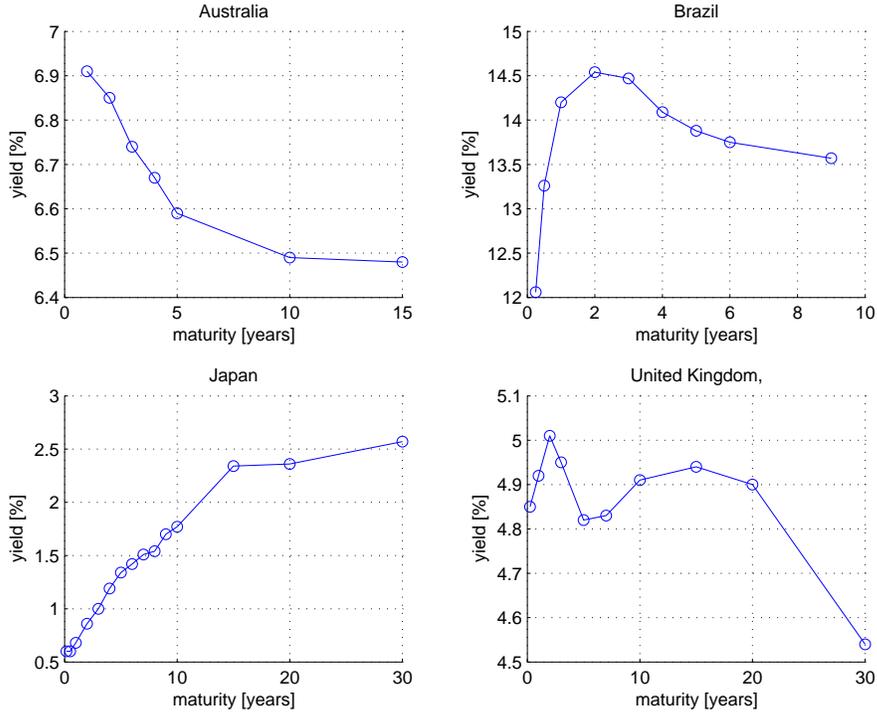} 
\caption{Examples of yield curves of governmental bonds: Australia, Brazil, Japan, United Kingdom 
(27th May 2008). 
Source: http://www.bloomberg.com}
\label{graf-vynosy}
\end{center}
\end{figure}

The term  structure of bond prices (or yields) is a function of time to maturity, state variables like e.g. instantaneous interest rate as well as several model parameters. It describes a functional dependence between the time to maturity of a discount bond and its present price. The yield of bonds, as a function of maturity, forms a term structure of interest rates. Figure~\ref{graf-vynosy} shows the different shapes of term structures observed on the market based on data by Bloomberg. We can observe various type of functional dependences including non-monotone term structures (Brazil) or term structures having two local maxima (UK). Interest rate models are often formulated in terms of stochastic differential equations (SDE) for the instantaneous interest rate (or short rate) as well as equations for other relevant quantities like e.g. volatility of the short rate process. In one-factor models there is a single
stochastic differential equation for the short rate. The volatility of
the short rate process is given in a deterministic way. They can be written in the form
\begin{equation}
dr= \mu(t,r) dt + \sigma(t,r) dw,
\label{dr-1f}
\end{equation}
where $w$ is a Wiener process. We recall that a stochastic process $\{ w(t), t \geq 0 \}$  is called a Wiener process if $w(0)=0$, every increment $w(t+\Delta t)-w(t)$ has the normal distribution $N(0,\Delta t)$, the increments $w(t_n)-w(t_{n-1})$, $w(t_{n-1})-w(t_{n-2})$, $\dots$, $w(t_2)-w(t_1)$ for $0 \leq t_1 < \dots < t_n$   are independent and paths of the process are continuous (see e.g. \cite{kwok}). The function $\mu$ in (\ref{dr-1f}) determines the trend in evolution of the short rate, function $\sigma$ the nature of stochastic fluctuations. The price of a discount bond $P(t,r)$ at time $t$ when the value of short rate is $r$, is known to be a solution of the partial differential equation
\begin{equation}
\frac{\partial P}{\partial t} + (\mu(t,r) - \lambda \sigma(t,r)) \frac{\partial P}{\partial r} + \frac{\sigma^2(t,r)}{2} \frac{\partial^{2}P}{\partial r^2} - rP = 0,
\label{pdr-1f}
\end{equation}
with the terminal condition $P(T,r)=1$ for any $r\ge 0$ where $T>0$ is a maturity of the bond. Here $\lambda$ stands for the so-called market price of risk. The above linear parabolic PDE is derived by constructing a riskless portfolio of bonds and using It\=o's lemma for evaluating differentials of a stochastic portfolio that is balanced by buying or selling bonds with different maturities. We refer the reader for a comprehensive overview of term structure modeling to the book by Kwok \cite[Chapter 7]{kwok} for details. 

We also remind ourselves that bond prices determine interest rates $R(t,r)$ by the formula $P=e^{-R (T-t)}$, i.e.
\[
R(t,r)=-\frac{1}{(T-t)} \log P(t,r).
\]
One of the first models of the class (\ref{dr-1f}) has been proposed by Old\v rich Va\v s\'\i\v cek in \cite{vasicek}. In this model, the short rate process is assumed to  follow a stochastic differential equation:
\begin{equation}
dr=\kappa(\theta-r)dt+ \sigma dw,
\label{vasicek-sde}
\end{equation}
where $\kappa, \theta, \sigma >0$ are positive constants. Here $\sigma>0$ stands for volatility of random fluctuations of the short rate process. Deterministic part of the process $\kappa(\theta-r)$ represents a mean reversion process with a limit $\theta$, referred to as long term interest rate.  The speed of reversion is given by the parameter $\kappa > 0$. In this model, for a constant market price of risk $\bar\lambda$, the corresponding PDE for bond prices 
\begin{equation}
\frac{\partial P}{\partial t} + (\kappa(\theta-r)  - \bar\lambda) \frac{\partial P}{\partial r} + \frac{\sigma^2}{2} \frac{\partial^{2}P}{\partial r^2} - rP = 0
\label{pdr-vasicek}
\end{equation}
has an explicit solution $P(t,r)$ satisfying the terminal condition $P(T,r)=1$ for any $r\ge 0$. It has the form
\begin{equation}
P(t,r)=A(t) e^{-B(t) r},
\label{tvar-riesenia}
\end{equation}
where the functions $A$ and $B$ can be expressed in a closed form (see, e.g.  \cite{vasicek} or \cite{kwok}):
$$B(t)=\frac{1-e^{-\kappa(T-t)}}{\kappa}, \; \ln A(t)= (B(t)-(T-t))(\theta-\frac{\bar\lambda}{\kappa} - 
\frac{\sigma^2}{2 \kappa^2} ) - \frac{\sigma^2}{4 \kappa} B(t)^2. $$

There is a rich variety of several other models in which the SDE for the short rate is given by a general process of the form:
\begin{equation}
dr=(a+ b r)dt + \sigma r^{\gamma} dw.
\label{ckls}
\end{equation}
This class of short rate models includes the well-known Cox-Ingersoll-Ross model \cite{cir} with $\gamma=1/2$. A thorough comparison of these models is a topic of the paper by Chan, Karolyi, Longstaff and Sanders \cite{ckls}. Using generalized method of moments they estimated the model (\ref{ckls}) and they studied restrictions on parameters imposed in this models. Their result that the optimal value of the parameter $\gamma$ is approximately $3/2$ (which is more than previous models assumed), started a broad discussion on the correct
form of volatility. Let us note that their result is not universal, e.g. in \cite{gmm-libor}, using the same
estimation methodology but for LIBOR rates, $\gamma$ was estimated to be less than unity (which means that volatility is less than proportional to short rate, unlike in the result due to Chan, Karolyi, Longstaff and Sanders). Approximate formulae for bond prices when the short rate follows (\ref{ckls}) has been  developed recently by \cite{1f-approximation} and \cite{1f-approximation-2}.

\begin{figure}
\begin{center}
\includegraphics[width=0.6\textwidth]{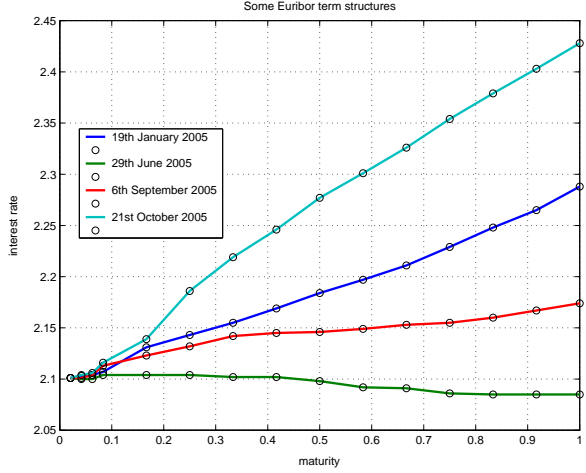}
\caption{Examples of real EURIBOR term structures. Source: http://www.euribor.org}
\label{real-term-structures-eu}
\end{center}
\end{figure}

In one-factor models, term structure of interest rates is a function of the short rate and model parameters. However, it means that as soon as  the parameters of the model are chosen, the term structure corresponding to a given short rate is uniquely determined. This is a simplification from the reality, as it can be seen in Fig. \ref{real-term-structures-eu}, showing the examples from EURIBOR data. To capture this feature, two-factor models are introduced. In the two-factor models there are two sources of uncertainty yielding different term structures for the same short rate. They may depend on the value of the other factor. Moreover, two-factor models have more variety of possible shapes of term structures.

%----------------------

A general two-factor model with the factors $x$, $y$ is given by the system of SDEs:
\begin{eqnarray}
dx&=& \mu_x dt + \sigma_x dw_1, \nonumber \\
dy&=& \mu_y dt + \sigma_y dw_2, \nonumber
\end{eqnarray}
where correlation between $dw_1$ and $dw_2$ is a assumed to be constant $\rho\in[-1,1]$, i.e. $E(dw_1 dw_2)=\rho dt$.  The short rate is a function of these two factors, i.e. $r=r(x,y)$. 

Let us denote by $P(t,x,y)$ the price of a zero coupon bond with maturity $T$, at the time $t$ when the values of the factors are $x$ and $y$.
the PDE satisfied by the bond price, which reads as (cf. \cite[Chapter 7]{kwok}) 
\begin{eqnarray}
\frac{\partial P}{\partial t} &+& (\mu_x - \lambda_x \sigma_x)\frac{\partial P}{\partial x} + (\mu_y - \lambda_y \sigma_y) \frac{\partial P}{\partial y}
\nonumber \\
&+& \frac{\sigma_x^2}{2} \frac{\partial^2 P}{\partial x^2} + \frac{\sigma_y^2}{2} \frac{\partial^2 P}{\partial y^2}
+
\rho \sigma_x \sigma_y \frac{\partial^2 P}{\partial x \partial y} - r(x,y) P =0.
\label{2f-general-pde}
\end{eqnarray}
The parameters $\lambda_x, \lambda_y$ stand for market prices of risk corresponding to factors $x$ and $y$, resp.  The equation for the bond price is equipped by the terminal condition at $t=T$, $P(T,x,y)=1$ for any $x$, $y$.

There are several ways of incorporating the second stochastic factor.
Based on experimental data from real financial market it is reasonable to make an assumption that the market changes the volatility of the underlying process for the short rate. More precisely, the volatility of the stochastic process for the short rate is stochastic as well. An empirical confirmation of such an assumption can be found e.g. in the recent paper by the authors \cite{kybernetika}. In the so-called two-factor models with a stochastic volatility we allow the volatility to have a stochastic behavior driven
by another stochastic differential equation. As an example of such a two-factors model one can consider the Fong--Vasicek model (cf. \cite{fv}) in which the stochastic volatility follows a mean reverting Bessel-square root process. Another possibility is to consider the generalized Fong--Vasicek model in which the drift function is no longer linear but it may have cubic like behavior having three distinct roots representing possible steady states for dispersion. By this we can model so-called volatility clustering phenomenon observed in real markets (see \cite{kybernetika} for details). Now, as a consequence of the multidimensional It\=o's lemma the corresponding equation for the bond price
is a linear parabolic equation in two space dimensions. These spatial dimensions
correspond to the short rate and volatility.

Let us consider the process a Bessel square root process with a general drift function $\alpha$, 
\begin{equation}
dy=\alpha(y)dt+ \omega\sqrt{y}dw.
\label{general-proc}
\end{equation}
It is well known that the density distribution of a stochastic process is a solution to the Focker-Planck partial differential equation (see \cite{goodman}). 
Recall that the cumulative distribution function $\tilde F=\tilde F(y,t)=Prob(y(t)<y|y(0)=y_0)$ of the process $y=y(t)$ satisfying (\ref{general-proc}) and starting almost surely from the initial datum $y_0$ can be obtained from a solution $\tilde F=\partial \tilde F/\partial y$ to the so-called Focker-Planck equation for the density function:
\begin{equation}
\frac{\partial \tilde f}{\partial t} = \frac{\omega^2}{2} \frac{\partial^2}{\partial y^2} ( y \tilde f) - \frac{\partial}{\partial y}(\alpha(y) \tilde f) , \quad  \tilde f(y,0)=\delta(y-y_0)\,.
\label{focker-planck}
\end{equation}
Here $\delta(y-y_0)$ denotes the Dirac delta function located at $y_0$. The limiting density $f(y)=\lim_{t\to\infty} \tilde f(y,t)$ of the process is therefore a stationary solution to the Focker-Planck equation (\ref{focker-planck}). A stationary solution satisfying $f(y)=0$ for $y\le0$ is therefore a solution to the differential equation  (see \cite{kybernetika})
\begin{equation}
 \frac{\omega^2}{2} \frac{\partial}{\partial y} (y f) - \alpha(y) f = 0.
\label{focker-planck-limit}
\end{equation}

Concerning structural assumption made on the drift function $\alpha:R\to R$ we shall henceforth assume the following hypothesis:
\[
(A)\qquad\qquad 
\alpha \ \ \hbox{is a } C^1 \ \hbox{function on } [0,\infty),\ \  \frac{2\alpha(0)}{v^2} >1,\ \ \limsup_{y\to\infty} \frac{\alpha(y)}{y} <0.
\]

Now it follows from the assumption (A) made on the drift function $\alpha$ and \cite[Lemma 2]{kybernetika} that the stationary Focker--Planck equation (\ref{focker-planck-limit}) has a solution $f$ that can be explicitly expressed as:
\[
f(y) = C y^{-1}\exp\left(\frac{2}{v^2} \int_1^y \frac{\alpha(\xi)}{\xi} d\xi\right) = C y^{\frac{2\alpha(0)}{v^2} -1} \exp\left(\frac{2}{v^2} \int_1^y \hat\alpha(\xi) d\xi\right)
\]
for $y>0$ and $f(y)=0$ for $y \leq 0$. Here $\hat\alpha(y)= (\alpha(y)-\alpha(0))/y$ and $C>0$ is a normalization constant such that $\int_0^\infty f(y) dy =1$. For example, if we consider a mean reverting Bessel square root  process for the stochastic dispersion $y$, i.e. the drift function is linear $\alpha(y)=\kappa_y(\theta_y -y)$ then the limiting density distribution function $f$ is the Gamma distribution function with shape parameters $2\kappa_y \theta_y/\omega^2$ and $2\kappa_y/\omega^2$ (see Kwok \cite{kwok}).

%-----------------------------------------------------------------------
%                 SECTION: Averaging with respect to volatility
%-----------------------------------------------------------------------
\section{Averaging with respect to stochastic volatility and relation to one-factor models}

Knowing the density distribution $f$ of the stochastic
volatility we are  able to perform averaging of the bond price and the term structure
with respect to volatility. Unlike the short rate which is known from the market data on daily
basis, the volatility of the short rate process is unknown. The exact value of the stochastic volatility is not observable on the  market,  we can just observe its  statistical properties. 
Therefore such a volatility averaging is of special importance for practitioners.

We shall consider the following model with stochastic volatility:
\begin{eqnarray}
dr&=&\kappa(\theta-r)dt + \sqrt{y} dw_r, \\
dy&=&\alpha(y)dt+\omega \sqrt{y} dw_y, \label{sde-2f-model}
\end{eqnarray}
with uncorrelated increments $dw_r$ and $dw_y$ of a  Wiener process. The market prices of risk are assumed to have a form  $\lambda_r(r,y)=\lambda \sqrt{y}$ and $\lambda_y=\frac{\tilde{\lambda}}{\omega} \sqrt{y}$. Then the bond price $\pi(t,r,y)$ satisfies the following PDE:
\begin{eqnarray}
&&\frac{\partial \pi}{\partial t} + (\kappa(\theta-r) - \lambda y)\frac{\partial \pi}{\partial r} + (\alpha(y) - \tilde{\lambda} y) \frac{\partial \pi}{\partial y} + \frac{1}{2} y \frac{\partial^2 \pi }{\partial r^2} + \frac{\omega^2}{2} y \frac{\partial^2 \pi }{\partial y^2} - r \pi =0
\end{eqnarray}
with the terminal condition $\pi(T,r,y)=1$ for any $r,y\ge 0$. The explicit solution can be written as
\begin{equation}
\pi(t,r,y)=A(t,y) e^{-B(\tau)r}
\end{equation}
with the terminal conditions $A(y,T)=1$ for any $y>0$ and $B(T)=0$. The solution can be obtained by solving the following differential equations for the functions $A$ and $B$:
\begin{eqnarray}
&&-B^\prime + \kappa B-1=0, \label{B-eq}\\
&&\frac{\partial A}{\partial t} - B(\kappa \theta - \lambda y - \frac{y}{2}B)A + (\alpha(y)-\tilde{\lambda} y)\frac{\partial A}{\partial y}+ \frac{\omega^2}{2}y \frac{\partial^2 A}{\partial y^2} =0. \label{A-eq}
\end{eqnarray}

In what follows, we shall denote by $\langle\psi\rangle$ the averaged value of the function $\psi~: [0,\infty)\to R$ with respect to the limiting density $f$, i.e. $\langle\psi\rangle = \int_0^\infty \psi(y)f(y)\,dy$, where $f(y)$ satisfies the stationary Focker--Planck equation (\ref{focker-planck-limit})
The averaged bond price with respect to the limiting distribution of the volatility is given by
\begin{equation}
P(t,r)=\langle \pi(t,r,.) \rangle = a(t) e^{-B(t)r},
\end{equation}
where $a(t)=\int_0^{\infty} A(t,y) f(y) dy$.

The function $P=P(t,r)$ is a function of time $t$ and short rate $r$. Notice that it is the same functional dependence as for the bond price in one factor models, including, in particular, a solution to (\ref{vasicek-sde}) given by (\ref{tvar-riesenia}). However, we show that there is no such one-factor model yielding the same bond prices as those of  averaged bond prices $P(t,r)$  given by the averaging of the  two-factor model.

Now we are in a position to state our main result of the paper. We are going to prove that there is no one-factor interest rate model for the corresponding bond prices that is identical with the volatility averaged bond price $P(t,r)$ for any $t\in[0,T]$ and $r\ge0$. 
Suppose to the contrary that $P=P(t,r)$ is a bond price from one-factor model in which the short rate is assumed to follow a general SDE
\begin{equation}
dr=\mu(r)dt+\Omega(r)dw.
\label{general-proc2}
\end{equation}
Then is satisfies the PDE
\begin{equation}
\frac{\partial P}{\partial t} + (\mu(r) -\Lambda(r))\frac{\partial P}{\partial r} + \frac{1}{2} \Omega^2(r) \frac{\partial^2 P}{\partial r^2} - rP=0,
\label{general-pde}
\end{equation}
where $\mu(r)$ is the drift of the short rate process, $\Omega(r)$ is its volatility and $\Lambda(r)$ is the product of the volatility and the corresponding  market price of risk of the model. Substituting the form of the solution we obtain  that
\begin{equation}
\frac{a^\prime(t)}{a(t)B(t)} = \kappa r + \mu(r) - \Lambda(r) - \frac{1}{2} \Omega^2(r) B(t).
\end{equation}
We see that the left hand side is a function of $t$ only. We denote it by $\phi(t)$, i.e. $\phi(t)=\dot{a}(t)/(a(t)B(t))$. Then,
\begin{equation}
\kappa r + \mu(r) - \Lambda(r) = \phi(t) + \frac{1}{2} \Omega^2(r) B(t),
\end{equation}
and so the right hand side is constant with respect to $t$. Hence for any $t$ it equals $\phi(T) + \frac{1}{2} \Omega^2(r) B(T) = \phi(T)$, which is a constant denoted by $K$. We have
\begin{equation}
\kappa r + \mu(r) - \Lambda(r) = \frac{1}{2}  \Omega^2(r)B(t)+\phi(t) = K,
\end{equation}
from which it follows that $\Omega(r)\equiv \bar\Omega$ is a constant and that $\mu(r)-\Lambda(r)=K-\kappa r$. 

Let us denote by $\sigma^2$ and $d$  the first two statistical moments of the random variable with respect to the limiting density function $f$, i.e.
\begin{eqnarray}
\sigma^2&=&\langle y \rangle = \int_0^{\infty} y f(y) dy, \quad 
d = \langle y^2 \rangle = \int_0^{\infty} y^2 f(y) dy. \nonumber
\end{eqnarray}

We know that $a(T)=1$ and from the expression
\begin{equation}
a'(t)=\left(K-\frac{\bar\Omega^2}{2}B(t)\right)a(t)B(t)
\end{equation}
we can recursively compute the values of the time derivatives of function $a$ at time $T$:
\begin{eqnarray}
a'(T)&=&0, \label{eq-1} \\
 a''(T)&=&-K, \label{eq-2}  \\
 a'''(T)&=&-K \kappa -\bar\Omega^2, \label{eq-3}  \\
a''''(T) &=& 3 K^2 - 3\bar\Omega^2 \kappa -K\kappa . \label{eq-4} 
\end{eqnarray}

\medskip
Another way of computing these derivatives is using the expression 
\[
a(t)=\int_0^{\infty}A(t,y)f(y)dy
\] 
and the partial differential equation (\ref{A-eq}) for the function $A$ and the stationary Focker--Planck equation for the limiting density function $f$. Indeed, by integration by parts and taking into account boundary conditions $yf(y)=0$ for $y=0,+\infty$ for the limiting density $f$ we obtain
\begin{eqnarray}
&&\int_0^\infty \left(\alpha(y) \frac{\partial A}{\partial y} + \frac{\omega^2}{2}y \frac{\partial^2 A}{\partial y^2}\right) f(y)  dy \nonumber \\
&&= \int_0^\infty \left(- \frac{\partial }{\partial y}(\alpha(y) f(y))  + \frac{\omega^2}{2} \frac{\partial^2 }{\partial y^2}(yf(y)) \right) A dy = 0.\nonumber
\end{eqnarray}
Furthermore, by (\ref{focker-planck-limit}), we have
\[
\int_0^\infty \frac{\partial A}{\partial y} y f(y) dy 
= - \frac{2}{\omega^2} \int_0^\infty A \alpha(y) f(y) dy.
\]
Therefore
\[
a^\prime(t) = \int_0^\infty \left(B(t)\left(\kappa\theta - \lambda y -\frac{y}{2} B(t)\right) - \frac{2\tilde\lambda}{\omega^2}\alpha(y) \right) A(t,y) f(y) dy.
\]
Now taking into account PDE (\ref{A-eq}) for the function $A(t,y)$ we can recurrently evaluate
\[
A(T,y)=1,\ 
\frac{\partial A}{\partial t}(T,y) = 0,\ 
\frac{\partial^2 A}{\partial t^2}(T,y) =  -\kappa\theta +\lambda y,
\]
\[
\frac{\partial^3 A}{\partial t^3}(T,y) =  -\kappa^2\theta  - (1-\kappa\lambda) y -\lambda(\alpha(y)-\tilde\lambda y),\ 
\]
for any $y>0$. Using the above expressions and the identities $B(T)=0, B^\prime(T) = -1, B^{\prime\prime}(T)=-\kappa,  B^{\prime\prime\prime}(T)=-\kappa^2$, after straightforward computations we obtain 
\begin{eqnarray}
a'(T)&=&0, \label{eq-5} \\
 a''(T)&=&-\kappa \theta + \lambda  \sigma^2 \label{eq-6}
\end{eqnarray}
and by comparing (\ref{eq-2}) and (\ref{eq-6}) we obtain the expression for the constant $K$ in terms of the model parameters as 
\begin{equation}
K=\kappa \theta - \lambda  \sigma^2.
\end{equation}
Computing the next derivative we end up with
\begin{equation}
a'''(T)=\tilde{\lambda} \lambda  \sigma^2 - \kappa^2 \theta + \kappa \lambda  \sigma^2 - \sigma^2 \label{eq-7}
\end{equation}
and by comparing (\ref{eq-3}) and (\ref{eq-7}) we can express the volatility $\bar\Omega$ as 
\begin{equation}
\bar\Omega^2 = \sigma^2 (1 - \tilde{\lambda} \lambda ) .
\end{equation}
Notice that the PDE for the averaged bond price now reads as follows:
\[
\frac{\partial P}{\partial t} + (\kappa(\theta -r)-\lambda\sigma^2)\frac{\partial P}{\partial r} + \frac{\sigma^2(1 - \tilde{\lambda} \lambda) }{2} \frac{\partial^2 P}{\partial r^2} - rP=0,
\]
which is the PDE corresponding to the classical one-factor Vasicek interest rate model. Now we fully determined the drift function minus market price function $\mu(r)-\Lambda(r)$ as well as the volatility function $\Omega(r)$ in (\ref{general-pde}). 

In order to achieve contradiction we finally compute the fourth derivative as
\begin{eqnarray}
a''''(T) &=& 3 \lambda^2 d + (-6 \kappa \theta \lambda + \kappa^2 \lambda - 3 \kappa + \tilde{\lambda}(\kappa \lambda -1 + \lambda \tilde{\lambda}))  \sigma^2 \nonumber \\ 
&& + 3 \kappa^2 \theta^2 -\kappa^3 \theta + \frac{2}{\omega^2} \tilde{\lambda} \lambda \int_0^{\infty} \alpha^2(y)f(y) dy.\label{eq-8}
\end{eqnarray}
Comparing (\ref{eq-4}) and (\ref{eq-8}) we get the condition
\begin{equation}
 \sigma^2 (2 \kappa \tilde{\lambda} \lambda + 1 - \lambda \tilde{\lambda}^2) = \frac{2}{\omega^2} \tilde{\lambda} \lambda \int_0^{\infty} \alpha^2(y)f(y) dy + 3 \lambda^2 (d-\sigma^4).
\end{equation}
 However, the latter equality can not be satisfied for general choice of model parameters. Indeed, setting  $\lambda=0$, we obtain  $0=\sigma^2=\int_0^\infty y f(y) dy$ which is not possible as $f(y)>0$ for $y>0$. 

\bigskip
Summarizing, we have shown the following theorem:
\medskip
\begin{theorem}
Consider the generalized Fong-Vasicek two-factors model with stochastic volatility (\ref{sde-2f-model}) and the averaged bond price $P(t,r)$ with respect to the limiting distribution of the stochastic dispersion. Then there is no one-factor interest rate model (\ref{general-proc2}) with corresponding PDE for the bond price (\ref{general-pde}) yielding the same bond prices as the averaged values $P(t,r)$ from the two-factor model.
\end{theorem}

\section*{Acknowledgment}
The support from the grant VEGA 1/3767/06 is kindly acknowledged.


\begin{thebibliography}{10}

\bibitem{gmm-libor}
L.C.~Adkins and T.~Krenbiel: Mean reversion and volatility of short-term London Interbank Offered Rates: 
An empirical comparison of competing models. International Review of Economics and Finance {\bf 8}, 1997, 45-55.

\bibitem{brigo-mercurio}
D.~Brigo and F.~Mercurio: 
Interest rate models -- Theory and practice. With smile, inflation and credit. Springer 2006.


\bibitem{ckls}
K.C.~Chan, G.A.~Karolyi, F.A.~Longstaff and A.B.~Sanders: An Empirical Comparison of Alternative Models of the Short-Term Interest Rate. The Journal of Finance {\bf 47}, 1992, 1209-1227.

\bibitem{1f-approximation}
Y.~Choi and  T.S.~Wirjanto: An analytic approximation formula for pricing zero coupon bonds. 
Finance Research Letters {\bf 4}(2), 2007, 116-126.

\bibitem{cir}
J.C.~Cox, J.E.~Ingersoll and S.A.~Ross: A Theory of the Term Structure of Interest Rates. Econometrica 
{\bf 53}, 1985, 385-408.


\bibitem{fv}
H.G.~Fong and  O.A.~Vasicek: Fixed-Income Volatility Management. Journal of Portfolio Management, 
1991, 41-46.

\bibitem{stoch-vol} J.-P.~Fouque, G.~Papanicolaou and K.R.~Sircar: 
Derivatives in Markets with Stochastic Volatility. Cambridge University Press 2000.

\bibitem{goodman}
J.~Goodman, K.S.~Moon, A.~Szepessy, R.~Tempone and G.~Zouraris: 
Stochastic and partial differential equations with adapted numerics. 
Royal Institute of Technology, Stockholm. www.math.kth.se/szepessy/sdepde.pdf

\bibitem {kwok}  
Y.K.~Kwok: Mathematical Models of Financial Derivatives. Springer--Verlag, Berlin 1998.

\bibitem{1f-approximation-2}
B.~Stehl\'ikov\'a and D. \v{S}ev\v{c}ovi\v{c}: Approximate formulae for pricing zero coupon bonds and their asymptotic analysis, to appear in: International J. Num. Anal. and Modeling.

\bibitem{herlany}
B.~Stehl\'ikov\'a: Averaged Bond Prices for Fong-Vasicek and the Generalized Vasicek Interest Rates Models. 
Proceeding of MMEI, Eds. K. Cechl\'arov\'a, M. Halick\'a, V. Borbe\v lov\'a, V. Lacko, 2007, 166-175.


\bibitem{iscam05}
B.~Stehl\'ikov\'a: Modeling Volatility Clusters with Application to Two-Factor Interest Rate Models. Journal of Electrical Engineering {\bf 56} (12/s) (2005), 90--93.

\bibitem{kybernetika}
B.~Stehl\'ikov\'a and D. \v{S}ev\v{c}ovi\v{c}:
On the singular limit of solutions to  the CIR interest rate model with stochastic volatility. 
Submitted.

\bibitem{vasicek} 
O.A.~Vasicek: An Equilibrium Characterization of the Term Structure. Journal of Financial Economics {\bf 5}, 1977, 177-188.

\end{thebibliography}
\end{document}